# Meta-study of laser power calibrations ranging 20 orders of magnitude with traceability to the kilogram*

*Paul A. Williams[1], Matthew T. Spidell[1], Joshua A. Hadler[1], Thomas Gerrits[1], Amanda Koepke[1], David Livigni[1], Michelle S. Stephens[1], Nathan A. Tomlin[1], Gordon A. Shaw[2], Jolene D. Splett[1], Igor Vayshenker[1], Malcolm G. White[1,3], Chris Yung[1], John H. Lehman[1]*

[1]National Institute of Standards and Technology, Boulder, CO 80305, USA
[1]National Institute of Standards and Technology, Gaithersburg, MD 20899, USA
[3]Department of Physics, University of Colorado, Boulder, CO 80309, USA

**Abstract**

Laser power metrology at the National Institute of Standards and Technology (NIST) ranges 20 orders of magnitude from photon-counting ($10^3$ photons/s) to 100 kW ($10^{23}$ photons/s at a wavelength of 1070 nm). As a part of routine practices, we perform internal (unpublished) comparisons between our various power meters to verify correct operation. Here we use the results of these intercomparisons to demonstrate an unbroken chain tracing each power meter's calibration factor to the NIST cryogenic radiometer (our lowest uncertainty standard, whose SI traceability is established through the volt and ohm units). This yields the expected result that all the NIST primary standard measurement techniques agree with each other to within their measurement uncertainty. Then, these intercomparison results are re-mapped to describe the agreement of the various techniques with our radiation-pressure-based power measurement approach, whose SI traceability is established through the kilogram. Again, agreement is demonstrated to within the measurement uncertainty. This agreement is reassuring because the measurements are compared with two entirely different traceability paths and show expected agreement in each case. The ramifications of this agreement as well as potential means to improve on it are discussed.

We demonstrate SI measurement traceability of our single-photon power measurement through the kilogram with less than 3 % relative expanded uncertainty (obtained for a coverage factor $k$=2 defining an interval having a level of confidence of approximately 95 %).

## I.    Introduction

The diverse applications of lasers brings an equally expansive range of laser power levels. In the attowatt regime, single-photon detectors are employed for metrology phase estimation [1], Bell tests [2, 3], exotic quantum states of light, low-light imaging and ranging [4, 5], and quantum key distribution [6, 7]. Laser range finding, and target designation can require measurement of peak power levels on the order of picowatts. Fiber optic telecommunications involve powers ranging from nanowatts to milliwatts. Medical applications use laser power levels from microwatts to watts. Laser processing such as annealing, cutting, welding, and additive manufacturing involve power levels from tens of watts to many kilowatts. Extreme applications such as laser drilling for oil and gas, advanced materials testing and aerospace thermal testing range from tens of kilowatts to the order of one-hundred kilowatts [8] and above.

From 1974 to the present, NIST has maintained a high-accuracy laser power metrology capability, which currently spans 20 orders of magnitude comprising several primary and secondary standard measurement techniques. A primary standard measurement of laser power is one that measures laser power without reference to other laser power measurement techniques [9]. Secondary (transfer) standard measurement techniques on the other hand must be calibrated by primary standards [9, 10]. Table I describes the various measurement approaches we use and their coverage ranges.

| Standard type | Laser power range | Wavelength range (nm) | Nomenclature | Mechanism | Relative expanded uncertainty ($U_{\hat{\mathcal{P}}_a}$ for $k=2$) |
|---|---|---|---|---|---|
| Primary | 10 µW – 1 mW | 850, 1295, 1550 | Optical fiber cryogenic radiometer (OFCR) | Planar absorbing bolometer [11] | 0.1 - 0.4 % [11, 12]† |
| Primary | 100 µW – 1 mW | 458-1550 | Laser optimized cryogenic radiometer (LOCR) | Electrical substitution radiometer [13] | 0.02 - 0.05 % |
| Primary | 100 µW – 2 W | 325-2000 | Mid-power calorimeter (C-series) | Isoperibol calorimeter [14] | 0.86 % |
| Primary | 100 µW – 1 W | 325-2000 | Next Generation C* | Planar absorbing bolometer [15] | 0.3 % |
| Primary | 2 W – 10 kW | 800-10600 | High power calorimeter (K-series) | Isoperibol calorimeter [16] | 1.1 - 1.5 % |
| Primary | 1 kW – 10 kW | 1070, 10600 | Flowing water optical power meter (FWOPM) | Heat balance calorimeter [17] | 1.6 % |
| Primary | 1 kW – 50 kW** [18, 19] | 1070 | Radiation pressure power meter (RPPM) | Photon momentum [20] | 1.6 % |
| Secondary | 0.5 fW – 500 fW | 851.7, 1533.6, 1550 | Single photon avalanche photodiode (SPAD) | Avalanche photodiode [21-23] | 1.53 %†† |
| Secondary | 10 µW – 1 mW | 633-1650 | Optical fiber power meter (OFPM) | Electrically calibrated pyroelectric radiometer [24] | 0.32 - 0.44 %†† |

**Table I.** Laser power measurement services supported at the National Institute of Standards and Technology. The relative expanded uncertainty $U_{\hat{\mathcal{P}}_a}$ represents a coverage factor $k=2$ defining an interval having a level of confidence of approximately 95 %. †The published uncertainty for the optical fiber cryogenic radiometer (OFCR) is 0.4 %, but our most recent work has demonstrated a decreased uncertainty of ~ 0.1 %. ††Relative expanded uncertainty of secondary standards is reported from traceability to the laser optimized cryogenic radiometer (LOCR). *The mid-power bolometer "Next Generation C" is still under development but is included here for completeness. **For the radiation pressure power meter (RPPM) we list the upper power range as 50 kW only because it is the highest power for which it has been rigorously tested – we expect full operability at higher powers.

Not only are significant differences in laser power measurement technology needed over such a large range, but traceability to the International System of Units (SI) is required. These diverse technologies must all achieve optical watt traceability to the defining constants through a combination of base and derived units consistent with the 2018 redefinition of the SI [25]. The path or combination of methods used to establish this traceability is important in terms of the accumulated uncertainty and is illustrated in Figure 1. Currently, all but one of our primary standard laser power measurement techniques use thermal comparators where optical heating is compared to that delivered by an 'electrically represented watt' traceable to the defining constants through the volt, ohm (and second in the case of calorimeters) [14, 16]. Alternatively, at the highest continuous wave (CW) laser powers, we also employ a radiation-pressure-based laser power meter with a significantly different traceability chain. This device is also a comparator, but instead one for which force delivered by optical power is compared to the force generated by a calibrated test mass. This results in traceability through the kilogram, the meter, and the second [20].

One purpose of this paper is to illustrate the level of agreement between laser power measurement techniques which are traceable through the volt, ohm, (and second in the case of calorimeters) to measurements that are traceable through the kilogram, meter, and second. This has the obvious benefit of further establishing the validity of our measurements through independent means. But, it also has a more forward-looking application in terms of the quantum realization of the SI. As will be discussed, radiation pressure-based measurement of laser power offers a means to realize the optical watt directly through force metrology, potentially producing uncertainty improvements for laser power radiometry as a whole. The mapping described here indicates the current uncertainties with which all of our laser power measurements can be made traceable through the kilogram and highlights areas for uncertainty reduction.

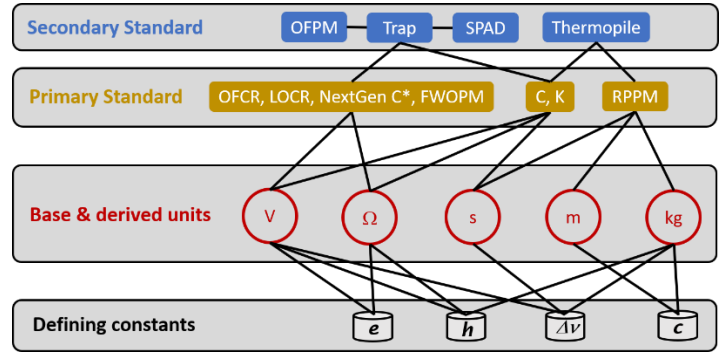

**Figure 1.** Traceability map for the NIST primary and secondary laser power measurement standards. The secondary (transfer) standards are abbreviated according to Table I as "OFPM" (optical fiber power meter), "Trap" (silicon/germanium trap-based photodiode detectors), SPAD (single-photon avalanche photodiode), and "Thermopile" (thermopile based thermal detector). The primary standards are as described in Table I, "OFCR" (optical fiber cryogenic radiometer), "LOCR" (laser-optimized cryogenic radiometer), "C" (mid-power range isoperibol calorimeter), "NextGenC" (the next generation of C-series power meter), "FWOPM" (flowing water optical power meter), "K" (high-power range isoperibol calorimeter), and RPPM (radiation pressure power meter). The relevant defining constants are $e$ (electronic charge), $h$ (Planck's constant), $\Delta \nu$ (cesium hyperfine splitting frequency), and $c$ (the speed of light in vacuum). *At the time of publication, the NextGenC is not yet fully validated as a primary standard but is included for completeness.

This work was motivated by technical progress expanding the upper and lower bounds of our measurement service capability. Recently we extended our low power measurement capability to include the single-photon-detection regime [26]. At the opposite end of the power range, we have used radiation pressure to measure the power of CW lasers approaching 100 kW [18-20]. Our lowest-uncertainty system, which is based upon a cryogenic cavity radiometer (LOCR), is now in the process of being replaced by a planar-absorber-based optical fiber cryogenic radiometer (OFCR) [11]. This minimizes the traceability chain and will eventually lead to lower uncertainties in optical fiber power measurement.

In Section II, the coverage range of CW laser power metrology at NIST is described. In Section III, the process of intercomparing the various power meters is explained and the results are presented. Section IV discusses the ramifications of the various levels of agreement. This includes suggesting means to further reduce measurement uncertainty for particular power ranges and ways to improve traceability under the redefined SI. Appendix A provides background, presenting a summary of the operation of each of our power measurement techniques, their range, and traceability path to the SI. Appendix B gives a mathematical description for radiation-pressure-based power measurements of the relationship between mass and optical power to the defining constants in the SI. Appendix C provides a description of the basis for our uncertainty expressions related to the comparison.

## II. Range of measurement capability

Figure 2 shows the power and uncertainty range typically covered by the various laser power measurement techniques at NIST. The uncertainty assignments (Figure 2, vertical axis) are taken from the lower end of the values expressed in Table I and represent the relative expanded uncertainties (obtained for a coverage factor $k=2$ [27] defining an interval having a level of confidence of approximately 95 %) [28].

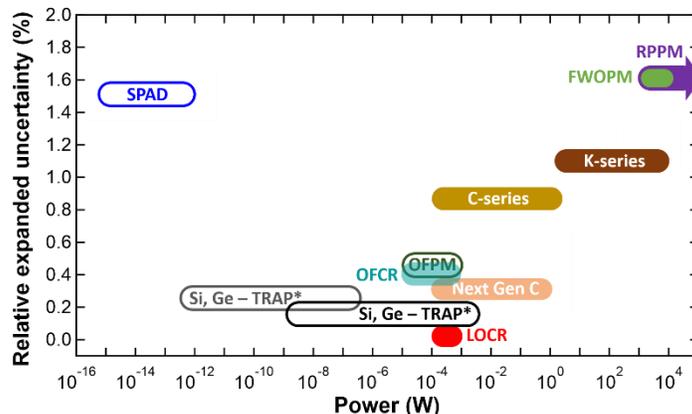

**Figure 2.** Expanded uncertainty for the various laser power measurement techniques in use at NIST plotted as a function of laser power measurement range. Solid symbols denote primary standard techniques and open symbols represent secondary (transfer) standard techniques. *Si and Ge trap detectors experience dramatic extensions of their coverage range through linearity characterization. These measurement techniques are described in Table I and Appendix A. The arrow on the end of RPPM symbol indicates no upper power limit has yet been tested.

For the primary standards, which are typically immobile, slow, or suitable for only a narrow range of injected power we supplement our power measurement capabilities with secondary standards having greater portability and/or dynamic range but no intrinsic traceability to the SI. Secondary standards require calibration by a primary standard. Their range of power coverage can be extended beyond their primary standard calibration points by measuring the linearity of their response [24, 29, 30].

### III. Measurement intercomparisons

Each of our primary standard laser power measurement techniques derive their SI traceability through the measurement of quantities other than laser power. For example, the thermal-based electrical substitution devices are traceable through measurements of voltage and resistance (and time in the case of calorimeters). Therefore, the measurement uncertainty of a primary standard is established without reference to any other laser power measurement technique. However, we do periodically compare our primary standard techniques to each other in order to identify unexpected changes in performance or equipment failure. To directly compare two primary standard power meters, they would have to be portable to operate in the same location and have overlapping power and wavelength ranges. This is the case only for the K-Series, FWOPM, and RPPM power meters. For the rest of our power meters, comparisons have been carried out by comparing one primary standard with a secondary (transfer) standard, and then comparing that transfer standard to another primary standard.

### III.a. Comparison Factor

We can quantify agreement between any two laser power meters (denoted generally as "$a$" and "$b$") through a measurement comparison where each meter measures the same nominal laser power (normalized against power drift between measurements). A comparison factor $K_{a,b}$ is defined such that if each power meter were illuminated by the same laser power $P_0$, the comparison factor gives the ratio of laser power reported by the two meters,

$$K_{a,b}(P_0) = \frac{P_a(P_0)}{P_b(P_0)}, \tag{1}$$

where $P_a(P_0)$ and $P_b(P_0)$ represent the power measured by meters "*a*" and "*b*" respectively when illuminated with a true power $P_0$. The practical use of this comparison factor is to represent how the power measured by meter "*a*" must be corrected to agree with that measured by the meter "*b*" for any power $P$ within the range over which $K_{a,b}$ is valid

$$P_b(P) = \frac{P_a(P)}{K_{a,b}(P_0)}. \tag{2}$$

The uncertainty of the comparison factor describes the uncertainty on the agreement of the two measurements.

In the special case where meter "*b*" is a primary standard and meter "*a*" is the device under test (DUT), $K_{a,b}$ becomes the calibration factor for power meter "*a*". If the two meters are in perfect agreement with each other, $K_{a,b}$ will be unity. But, $K_{a,b}$ alone gives no information on the individual accuracy of meter "*a*" or "*b*". The role of meter "*a*" and meter "*b*" can be interchanged by knowing that $K_{a,b} = (K_{b,a})^{-1}$.

This comparison factor can be generalized to include the case when meters "*a*" and "*b*" cannot be collocated for comparison or have non-overlapping operating ranges of wavelength and/or power. In that case, these meters cannot be compared directly and a third (intermediate) power meter must be used. This intermediate secondary (transfer) standard is compared to power meters "*a*" and "*b*" independently and the results can be combined to yield a comparison factor between "*a*" and "*b*" even without direct comparison between the two.

This can be done as follows. Suppose power meter "*a*" is able to measure incident light of power $P_1$ and reports $P_a(P_1)$. For a power $P_2$ that is out of the range of meter "*a*", we can extrapolate to say that if meter "*a*" could measure at the level of $P_2$, we would expect it to report a power scaled to its measurement of $P_1$

$$P_a(P_2) = P_a(P_1)\frac{P_2}{P_1}. \tag{3}$$

Further, suppose power meter "*b*" cannot measure at a power level of $P_1$, but can measure at $P_2$, reporting a power $P_b(P_2)$. Equations (1) and (3) can be combined to yield the definition of a (virtual) comparison factor

$$K_{a,b}(P_2) = \frac{P_a(P_2)}{P_b(P_2)} = \frac{P_a(P_1)}{P_b(P_2)}\frac{P_2}{P_1}. \tag{4}$$

Of course, the true values of power levels $P_1$ and $P_2$ are not known. So, we enable this virtual comparison by introducing a transfer standard power meter "*x*" which can measure at both power levels $P_1$ and $P_2$. From equation (1) meter "*x*" can be compared to meter "*a*" at power $P_1$,

$$P_a(P_1) = \frac{P_x(P_1)}{K_{x,a}(P_1)}, \tag{5}$$

and compared to meter "*b*" at power $P_2$

$$P_b(P_2) = \frac{P_x(P_2)}{K_{x,b}(P_2)}. \tag{6}$$

Combining Equations (4-6) produces

$$K_{a,b} = \frac{K_{x,b}(P_2)}{K_{x,a}(P_1)}\gamma_x, \tag{7}$$

where

$$\gamma_x = \frac{P_x(P_1)}{P_1} \Big/ \frac{P_x(P_2)}{P_2} \qquad (8)$$

represents the power nonlinearity of power meter "$x$", and $\gamma_x = 1$ when the meter is perfectly linear. Similarly, we define a spectral responsivity ratio $\eta_x$ for the transfer meter "$x$"

$$\eta_x = \frac{P_x(P, \lambda_1)}{P_x(P, \lambda_2)}, \qquad (9)$$

where $P_x(P,\lambda)$ is the power reported by meter "$x$" for incident laser light of power P and wavelength $\lambda$. This permits a general expression for the comparison factor for power meters "$a$" and "$b$" with non-overlapping power and wavelength ranges

$$K_{a,b} = \frac{K_{x,b}(P_2,\lambda_2)}{K_{x,a}(P_1,\lambda_1)} \gamma_x \eta_x. \qquad (10)$$

The uncertainty on $\gamma_x$ and $\eta_x$ must be included in the uncertainty of the comparison factor $K_{a,b}$. In both the specific case of Equation (1) or the most general case of Equation (10), the comparison factor is a ratio. Its uncertainty indicates only the repeatability of the ratio of the powers reported by the two meters in question, not the full uncertainty of either power meter. Therefore, the uncertainty of $K_{a,b}$ denoted as $u_{K_{a,b}}$ comes only from the repeatability statistics of the comparative measurement and the uncertainties of $\gamma_x$ and $\eta_x$ if applicable.

A detailed explanation of the uncertainty expression for $K_{a,b}$ is given in Appendix C, but is summarized here. From Equation (C5), the uncertainty of $K_{a,b}$ is only statistical, depending on the relative standard uncertainties $u_{a,stat}$ and $u_{b,stat}$ of powers reported by power meters "$a$" and "$b$", respectively, as

$$U_{K_{a,b}} = 2\sqrt{u_{a,stat}^2 + u_{b,stat}^2}, \qquad (11)$$

where, as for all cases below, the capital '$U$' represents the relative expanded uncertainty, where a coverage factor $k=2$ defines an interval having a level of confidence of approximately 95 %. The subscript "*stat*" indicates the statistical component of uncertainty, often referred to as "Type A", and "*non-stat*" will indicate uncertainty components obtained through non-statistical means, often referred to as "Type B" [27] – these designations are not to be confused with the lower case "$a$" and "$b$" identifiers used here to differentiate the power meters. The uncertainty of $K_{a,b}$ does not indicate the individual uncertainty of either power meter, merely their agreement with each other.

The transitive relationship of Equation (10) allows any two power meters to be virtually compared if there is an unbroken chain of comparisons between them,

$$K_{n,1} = K_{n,n-1}K_{n-1,n-2} \ldots K_{3,2}K_{2,1} = \prod_{i=1}^{n-1} K_{i+1,i}. \qquad (12)$$

Since each of the *K* values in Equation (12) are measured independently of the others and have only statistical uncertainties, their errors are uncorrelated, and their covariance is zero. Therefore, the relative uncertainty of the net comparison factor (e.g. $K_{n,1}$ in Equation (12)) is just the quadrature sum of the relative statistical uncertainties of each of the direct comparisons, and the relative expanded uncertainty is

$$U_{K_{n,1}} = 2\sqrt{u_{K_{n,n-1}}^2 + u_{K_{n-1,n-2}}^2 + \cdots + u_{K_{2,1}}^2} = 2\sqrt{\sum_{i=1}^{n-1} u_{K_{i+1,i}}^2}. \qquad (13)$$

Among our primary standard power meters, the laser optimized cryogenic radiometer (LOCR) currently has the lowest measurement uncertainty and the radiation pressure power meter (RPPM) has a unique traceability path through the kilogram. We find it useful to compare our other power meters to these two primary standards using the comparison factor approach described above. Figure 3 illustrates the various comparisons we have made and the power levels and wavelengths at which they were carried out. Table II presents the results of these measurement comparisons performed between our various power meters and the associated uncertainties. Virtual comparison factors between power meters have been constructed by taking the product of the appropriate individual comparison factors.

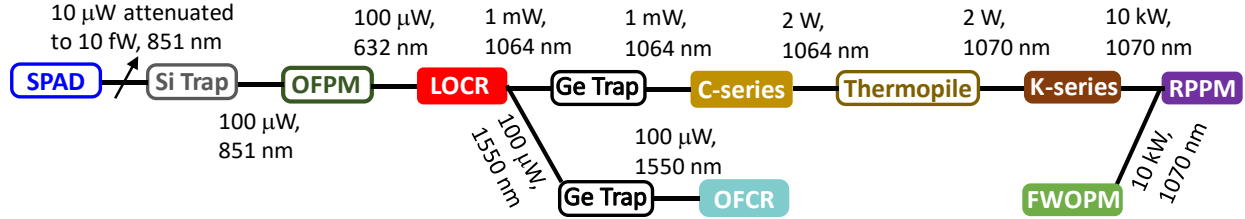

**Figure 3.** Illustration of the direct comparisons performed between the various power meters indicating the power and wavelength at which each was carried out. Solid-fill boxes denote primary standards, outline boxes denote secondary (transfer) standards. For the comparison between the Si Trap and SPAD, 10 μW of power was incident on the Si Trap and then attenuated by a calibrated 90 dB attenuator to yield 10 fW of incident power on the SPAD.

| DUT vs Standard | $K_{DUT,Std}$ | $U_{K_{DUT,Std}}$ | 1-$K_{DUT,Std}$ | $U_{\hat{\mathcal{P}}_{DUT \to Std}}$ |
|---|---|---|---|---|
| SPAD* vs LOCR | 1 | 1.53 % | 0 | 1.53 % |
| OFPM* vs LOCR | 1 | 0.44 % | 0 | 0.44 % |
| OFCR vs LOCR | 0.9985 | 0.38 % | 0.0015 | 0.38 % |
| C vs LOCR | 1.0030 | 0.90 % | -0.0030 | 0.90 % |
| K vs LOCR | 1.0064 | 1.3 % | -0.0064 | 1.3 % |
| RPPM vs LOCR | 1.0123 | 2.0 % | -0.0123 | 2.0 % |
| FWOPM vs LOCR | 1.0025 | 3.0 % | -0.0025 | 3.0 % |
| SPAD* vs RPPM | 1 | 2.6 % | 0 | 2.6 % |
| LOCR vs RPPM | 0.9878 | 2.0 % | 0.0122 | 2.1 % |
| OFPM* vs RPPM | 1 | 2.1 % | 0 | 2.2 % |
| OFCR vs RPPM | 0.9864 | 2.1 % | 0.0136 | 2.2 % |
| C vs RPPM | 0.9908 | 1.8 % | 0.0092 | 1.9 % |
| K vs RPPM | 0.9922 | 1.6 % | 0.0078 | 1.7 % |
| FWOPM vs RPPM | 0.9935 | 2.2 % | 0.0065 | 2.3 % |

**Table II.** Comparison results between various laser power meters at NIST. The top seven entries represent comparisons of seven power meters to LOCR and the bottom seven entries are comparisons of those power meters with RPPM. The comparison is labeled as the device under test (DUT) vs standard (Std). The standard in this case is either the LOCR or the RPPM. $K_{DUT,Std}$ is the comparison factor between the DUT and the Standard, and $U_{K_{DUT,Std}}$ is its relative expanded uncertainty. The expression 1-$K_{DUT,Std}$ denotes the disagreement between the DUT and the standard, and $U_{\hat{\mathcal{P}}_{DUT \to Std}}$ is the relative expanded uncertainty of the DUT when the traceability path is through the standard. The differences here between $U_{K_{DUT,Std}}$ and $U_{\hat{\mathcal{P}}_{DUT \to Std}}$ are insignificant due to the dominance of statistical uncertainty sources. The power levels and wavelengths at which the direct comparisons underlying these results are made are shown in Figure 3. *Indicates a secondary standard, the rest are primary standards.

### III.b. Calibration of Secondary (Transfer) Standards

Most meters reported in Tables I and II are primary standards, meaning reported laser power is based on SI traceability to quantities other than optical power. The remaining meters in Tables I and II (or those which were used to support the results in Table II) are secondary (transfer) standards in the form of pyroelectrics [24], silicon or germanium photodiodes arranged in an optical trap configuration [31] or thermopiles [32, 33]. These secondary standards are traceable to the SI through calibration with one of NIST's primary standard laser power meters. Calibration involves the same procedure as measuring the comparison factor of Equations (1) and (2) except that meter "$b$" is a primary standard. The output of meter "$a$" is then rescaled (divided by $K_{a,b}$) so that it reports a value that (on average) agrees exactly with power meter "$b$". Were this calibrated secondary standard then re-compared to the primary standard used to establish it's SI traceability, we would expect to measure a comparison factor of 1 (neglecting the randomness of a particular measurement). This is why the comparison factor equals 1 for the secondary standards of Table II.

### III.c. Uncertainty Traceable to the SI

We consider a power meter "$a$" that has traceability through power meter "$b$" and typically, "$b$" would be the primary standard. The comparison factor $K_{a,b}$ between these power meters estimates their agreement, and its uncertainty $u_{K_{a,b}}$ quantifies the uncertainty of this agreement, but does not address the absolute accuracy of either technique. The relative expanded uncertainty of power meter "$a$" having traceability through power meter "$b$" is expressed as $U_{\hat{p}_{a \to b}}$ and can be found as follows. We include the statistical uncertainties of power meter "$a$" and the comparison factor $K_{a,b}$ with the non-statistical uncertainty of power meter "$b$" as discussed in Appendix C and Equation (C12)

$$U_{\hat{p}_{a \to b}} = 2u_{\hat{p}_{a \to b}} = 2\sqrt{u_{a,stat}^2 + u_{K_{a,b}}^2 + u_{b,non-stat}^2}. \tag{14}$$

The statistical uncertainty component of power meter "$b$" ($u_{b,stat}^2$) is already contained in $u_{K_{a,b}}$ as seen in Equation (11). As an example, we have used this approach to quantify the uncertainty of our single-photon avalanche photodiode (SPAD) power meter with a traceability chain through the LOCR where

$$K_{SPAD,LOCR} = K_{SPAD,SiTrap} K_{SiTrap,OFPM} K_{OFPM,LOCR}. \tag{15}$$

Since the SPAD is not a primary standard, it has been calibrated to the LOCR (forcing a unity value of $K_{SPAD,LOCR}$). The relative expanded uncertainty of the single photon power meter is determined according to Equations (11), (13), and (14) to be $U_{\hat{p}_{SPAD}} = 1.53\ \%$ with traceability through the LOCR.

Figure 4 illustrates the comparison factors and uncertainties for several of our power meters when traceability is obtained by comparison with our highest accuracy primary standard power meter, the LOCR. Of course, many of the meters in question are themselves primary standards with independently established uncertainties. However, by evaluating the uncertainty achieved through this alternate traceability path through LOCR, we can compare the measurement agreement as well as identify the approach which yields the lowest uncertainty. Three sets of error bars are included in Figure 4. The thick vertical error bars have no horizontal cap and illustrate the relative expanded uncertainty $U_{K_{a,LOCR}}$ of the comparison factor from Table II. This value can be used as will be described in Equation (16) to assess the validity of agreement between a given technique and LOCR. The wide horizontal bars indicate the relative expanded uncertainty $U_{\hat{p}_{a \to LOCR}}$ of each power meter when traceability to the SI is established through the LOCR. The narrower horizontal bars are the intrinsic uncertainty $U_{\hat{p}_a}$ of each power meter determined through its own traceability path as a primary standard (from Table I). As secondary standards, the SPAD and OFPM do not have an intrinsic uncertainty apart from traceability through the LOCR and so $U_{\hat{p}_a} = U_{\hat{p}_{a \to LOCR}}$. For the OFCR and C-calorimeter, the statistical uncertainties of the comparison with the LOCR make their intrinsic primary standard uncertainty of no significant advantage compared to traceability through the LOCR. On the other hand, the K-calorimeter, RPPM and FWOPM have intrinsic uncertainties that are better than what can currently be obtained through LOCR traceability due to the significant number of nodes in the traceability chain and the large statistical uncertainty levels of the RPPM and FWOPM measurements.

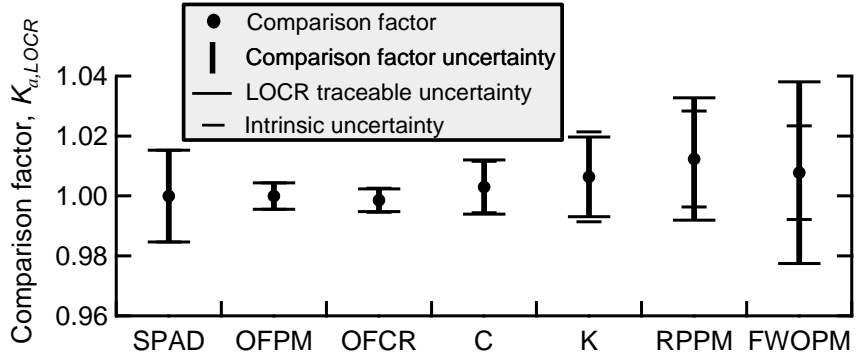

**Figure 4.** Comparison factor $K_{a,LOCR}$ (circles) indicating agreement between NIST's various optical power meters and the laser optimized cryogenic radiometer (LOCR). This plot contains three different error bars. The wide vertical error bars with no horizontal cap are the relative expanded uncertainty of the comparison factor $U_{K_{a,LOCR}}$. The wide horizontal bars indicate the relative expanded uncertainty of each power meter when traceability is through the LOCR $U_{\hat{\mathcal{P}}_{a \to LOCR}}$. As can be seen in Table II, for LOCR traceability, $U_{K_{a,LOCR}}$ is equal to $U_{\hat{\mathcal{P}}_{a \to LOCR}}$ to within two significant figures. The narrow horizontal error bars are the intrinsic relative expanded uncertainty of each power meter $U_{\hat{\mathcal{P}}_a}$ independent of its traceability through the LOCR and comes from Appendices A, and C. Note that since the SPAD and OFPM are not primary standards they have no intrinsic uncertainty apart from LOCR and so $U_{\hat{\mathcal{P}}_a} = U_{\hat{\mathcal{P}}_{a \to LOCR}}$ and for the OFCR and the C power meters, $U_{\hat{\mathcal{P}}_a}$ and $U_{\hat{\mathcal{P}}_{a \to LOCR}}$ are close enough to be nearly indistinguishable in this plot. All values and uncertainties are expressed as a fraction rather than percent.

The comparison factors and their uncertainties can be used to validate two power meter techniques as agreeing to within their combined measurement uncertainties. That is, when comparing power meter "$a$" with power meter "$b$", we have agreement between the two techniques within the uncertainty of the comparison when the following is true

$$|1 - K_{a,b}| \leq |U_{K_{a,b}}|, \qquad (16)$$

where $K_{a,b}$ is the appropriate comparison factor and $U_{K_{a,b}}$ is the relative expanded uncertainty of meter "$a$" with its traceability through meter "$b$" from Equations (11) and (13).

We can also map the traceability of these power meters through the RPPM for which traceability is established through the kilogram, meter, and second. The utility of such a path is that the potential sources for inequivalence are different from the volt, ohm and second based traceability path to the SI typically used, yet the inequivalence (1-$K_{DUT,RPPM}$) is well within the combined uncertainties. Figure 5 illustrates the comparison factor and uncertainties obtained in this manner. Note that we can measure power from the single-photon level to 50 kW and above, all traceable through the kilogram with a relative expanded uncertainty of less than 3 %. This is a strong validator of our measurement techniques in that calibration of each, through a completely different traceability path, yields an acceptable uncertainty.

Using Figure 5, it is instructive to compare the full uncertainty for this kilogram traceability $U_{\hat{\mathcal{P}}_{a \to RPPM}}$ to the uncertainty $U_{\hat{\mathcal{P}}_a}$ when traceability is through the DUT's intrinsic primary standard path. We find that traceability through RPPM currently adds significant uncertainty to most of the measurement techniques. Better uncertainties for kilogram traceability can be obtained through three approaches: reduction of the uncertainties associated with each comparison, reducing the number of comparisons needed, and/or by reducing the uncertainty of the RPPM itself which currently represents the largest contribution. The uncertainty for each of the RPPM comparisons includes the statistical uncertainties incurred through the comparison chain (~0.5 % for the comparisons with K-Series and ~0.8 % from comparisons through C-Series calorimeters). Reduction of these uncertainties would help. However, even more effective for kilogram traceability would be to allow traceability of a power meter through RPPM with fewer comparisons in between. This is achievable if the lower power limit of radiation pressure power measurements were reduced below a few watts. In that case, the RPPM could be directly compared to C-series without need to include the K-series (and its uncertainty) in the comparison chain. This might be done through the use of high accuracy attenuation [34]. However, with a current 1 kW lower limit for RPPM and a 2 W upper limit for the C-series measurement system, low uncertainty attenuation on the order of 30 dB would be needed which exceeds our current capability.

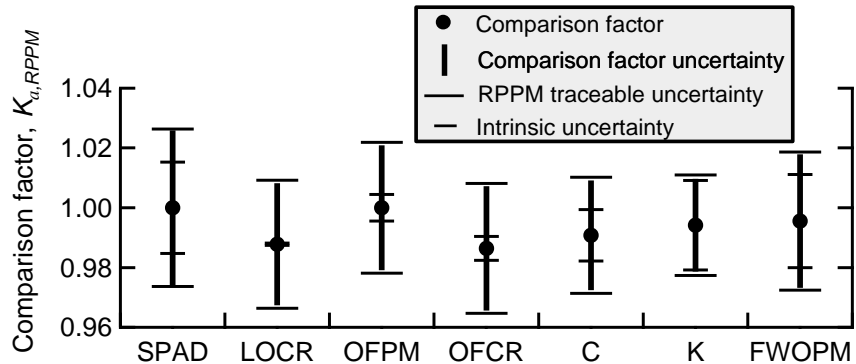

**Figure 5.** Comparison factor $K_{a,RPPM}$ (circles) indicating agreement between NIST's various optical power meters and the radiation pressure power meter (RPPM). This plot contains three different error bars. The thick vertical bars with no horizontal cap are the relative expanded uncertainty of the comparison factor $U_{K_{a,RPPM}}$. The wide horizontal bars indicate the relative expanded uncertainty of each power meter when traceability is through RPPM $U_{\hat{\mathcal{P}}_{a \to RPPM}}$. As can be seen in Table II, for RPPM traceability, $U_{K_{a,RPPM}}$ is close but not equal to $U_{\hat{\mathcal{P}}_{a \to RPPM}}$. The narrow horizontal error bars are the intrinsic relative expanded uncertainty $U_{\hat{\mathcal{P}}_a}$ of each power meter independent of its traceability through RPPM and comes from Appendices A, and C. All values and uncertainties are expressed as a fraction rather than a percent.

IV. **Discussion**

This intercomparison of NIST primary measurement systems for laser power demonstrates mutual agreement within their stated uncertainty, as illustrated in Figure 4. We demonstrate that our power meters can be made traceable through a cryogenic radiometer (the LOCR) without significantly increasing their uncertainty over their intrinsic value. Single-photon measurements can be made traceable through the kilogram with a mere 3 % relative expanded uncertainty accumulated across the 20 orders of magnitude power range. These uncertainties, incurred for the various power meters when traceability is established through the current radiation pressure technique (the RPPM) as quantified in Figure 5 are currently higher than what would be if traceability were established through a cryogenic radiometer (the LOCR) as quantified in Figure 4. This is due in large part to the higher measurement uncertainty of RPPM (1.6 %) compared to that of LOCR (0.02 % - 0.05 %).

Improvements in the sensitivity of radiation pressure power measurements would also enable measurements at lower optical powers. Table III lists historical demonstrations of radiation pressure-based power measurements. Many of these measurements were performed at the milliwatt level, but with varying degrees of measurement uncertainty and with instruments that are not well-suited for laser power metrology (difficult to calibrate or transport). We are currently developing portable radiation pressure-based power meters with improved sensitivities and lower minimum powers [35, 36]. However, direct measurements of radiation pressure at the lowest powers of NIST's current coverage range will be extremely challenging. For this task, cavity enhancement approaches are promising for amplifying the radiation pressure force for improved sensitivity [37, 38].

On a more fundamental level, the traceability chain of the radiation pressure approach to laser power measurement can be simplified. Currently, our radiation pressure power measurements require calibration of the force sensor using a mass standard [20]. This has traceability to the SI defining constants through the kilogram, meter, and second. Notably, knowledge of the local acceleration of gravity $g$ is required to complete such a comparison (relating force to mass). The motivation to relate optical power and the kilogram stems from two benefits. First, traceability through the kilogram allows calibration of optical power meters with an easily portable transfer standard – a reference mass. Second, the relationship between the force generated by radiation pressure and the force on a mass in a gravitational field can be used to improve measurement uncertainty for both high power laser calibration and small mass or force measurements.

| Sensor type | Pressure (Pa) | Agreement/ Uncertainty | Source / wavelength (nm) | Power /energy | Mirror dim. | Ref. (year) |
|---|---|---|---|---|---|---|
| Torsion | 8 – 13000 | 1 % /1 % | Lamp / NA | 100 mW | ~ 1 cm | [39, 40] (1903) |
| Torsion | < 0.013 | NA / 20 % | Lamp / NA | 77 mW | 0.4 mm | [41] (1901) |
| Torsion | < 0.007 | "order of magnitude" / NA | Ruby / 694.3 | 2.5 J | 2.5 cm | [42] (1964) |
| Torsion | 0.013 | NA / NA | N/A | 3 J | 1.4 cm | [43] (1962) |
| Torsion | < 0.007 | NA / 3 % | TEA $CO_2$ / 10600 | 2-6 J | 2 cm | [44] (1990) |
| Pendulum | Atm. | 20 % / 1 % | Nd:YAG / 1064 | 450 mW mod. | 3 mm | [45] (2014) |
| Pendulum | Atm. | 5 % / NA | HeNe / 632.8 | 7 mW | 15 mm | [46] (2009) |
| Cantilever | $6\times10^{-4}$ | NA / 2 % | SLD / 1555 | 6.5 mW mod. | 2 mm | [47] (2013) |
| Cantilever | Atm. | NA / NA | NA / 660 | 10 mW mod. | 10 μm | [48] (2015) |
| Circulating cavity | Atm. & vacuum | 21 % / NA | Diode / 1550 | 0.4 mW mod. | < 250 μm | [37] (2018) |
| Force balance | Atm. | 1.6 % / 1.7 % | Yb-fiber / 1070 | 1-50 kW CW | 150 mm | [18, 20] (2017,2018) |
| Force balance | Atm. | 5 % / 5 % | Raman fiber / 1363 | 1 W | N/A | [49] (2018) |

**Table III.** Published comparisons between optical power and radiation pressure force. "Pressure" is environmental pressure under which the measurement was made with "Atm." indicating the measurement was made at atmospheric pressure with no actual pressure value reported. "Agreement / Uncertainty" indicates the agreement between measured optical power and measured force values as reported by authors and the uncertainty reported by the authors. SLD is a superluminescent diode, and "mod." indicates that the applied power was intensity-modulated.

But, as described in Appendix B, a more minimal traceability path can be obtained since radiation pressure measurements of laser power are fundamentally a measure of force and not mass. Therefore, realization of the optical watt could be made by measuring the force of incident laser light using a Kibble balance (Watt balance) [50] or an Electrostatic Force Balance (EFB) [51, 52] both of which achieve SI traceability independent of the kilogram. This would eliminate the need to know $g$. The result would be an optical watt whose traceability would not include the kilogram. In fact, the force $F$ produced by light of power $P$ at normal incidence on a perfectly reflecting mirror is given simply in terms of the speed of light $c$ as [20]

$$F = \frac{2P}{c}. \tag{17}$$

This provides an elegant definition of the optical watt that no longer includes reference to a kilogram and relates force to optical power through a single fundamental constant – the speed of light $c$.

> *One watt of optical power is that which, upon normal reflection from a perfect mirror produces a force whose magnitude (in newtons) is equal to 2/c.*

Alternatively,

> *One newton of force is that which is produced when an optical power (in watts) of magnitude c/2 reflects normally from a perfectly reflecting mirror.*

We propose a low-uncertainty measurement of laser power using a modified Kibble balance to measure the force of radiation pressure without the need to reference the kilogram. The demonstrated uncertainty of the Kibble balance (which supported a determination of Planck's constant at the level of 13 parts per billion) [53] would allow for a simplified version to be used at the tens of kilowatt to one hundred kilowatt level for uncertainties that rival those currently achievable only through cryogenic radiometry.

## V. Conclusions

We have shown agreement better than 3 % between eight different measurement techniques spanning a power range 20 orders of magnitude with traceability through the kilogram via radiation pressure. Further expansion of this range is limited on the low end by the dark counts of our single photon detector. The upper power limit of our current radiation pressure power metrology has not yet demonstrated a constraint and we are pursuing testing opportunities for 100 kW lasers and above. This traceability through the kilogram has room for significant improvement in uncertainty as outlined above as well as via a path that does not require reference to the kilogram at all.

## Appendix A. Description of the various measurement methods

Each measurement technique will be described in terms of its operating principle, power and wavelength range, measurement uncertainty, and traceability path. All of these techniques except the radiation pressure power meter use a thermal approach to laser power measurement with electrical substitution to determine the input optical power. This is illustrated in Figure A1.

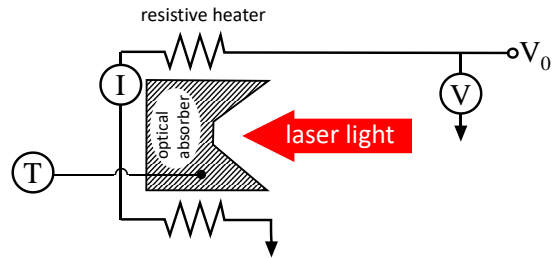

**Figure A1.** Electrical substitution approach to measuring input laser power. The optical absorber is instrumented with a thermometer (measuring temperature $T$), a current meter (measuring current $I$), and a voltmeter (measuring voltage $V$) Laser light incident on the optical absorber causes an increase in the absorber's temperature. Alternatively, a direct current (dc) voltage $V_0$ applied to the resistive heater can be used to cause an equal temperature rise for an accurately known electrical power $VI$. Measurements to characterize the inequivalence between optical and electrical heating allow an accurate measure of laser power as a function of absorber temperature.

### 1. Laser optimized cryogenic radiometer (100 μW – 1 mW, 458-1550 nm) – primary standard

Our lowest uncertainty measurements are carried out with a laser-optimized cryogenic radiometer (LOCR) [13] designed as a primary standard to measure laser power using the technique of electrical substitution [54], traceable through the NIST volt and ohm. The relative expanded uncertainty of the LOCR ranges from 0.02 % to 0.05 % depending on implementation.

The absorber cavity is heated with an electrical current while the temperature is monitored. When laser power is injected, the heater current is reduced to maintain a constant cavity temperature. The reduction in electrical power during the injection provides the measure of the injected laser power. This measurement is made traceable to the defining constants of the SI through a NIST-calibrated volt meter and shunt resistor.

2. **Optical fiber cryogenic radiometer (10 μW – 1 mW, 850 nm, 1295 nm, 1550 nm) – primary standard.**

The optical fiber cryogenic radiometer (OFCR) is NIST's cryogenic primary standard for optical fiber power measurement and calibration. It provides a direct traceability route for our secondary standard optical fiber power meter (OFPM). Two silicon micro-machined planar detectors, with vertically aligned carbon nanotube absorbers, thin film tungsten heaters and superconducting resistive transition edge temperature transducers, form the basis of the radiometer. Magnetic phase-change thermal filters ensure noise-free operation at 7.6 K. Light is coupled to the cryogenic detectors through an optical fiber. A fiber beamsplitter allows direct calibration of the device under test with the planar detectors. The system operates at a nominal radiant power level of 200 μW (−7 dBm). Traceability is through a NIST calibrated voltmeter and series resistor.

Measurement repeatability below 50 ppm is routinely achieved during a measurement cycle of 30 minutes. The expanded measurement uncertainty is currently 0.4% [11]. However, work in progress is reducing the uncertainty sources and we expect to soon achieve a relative expanded uncertainty of less than 0.1 % [12].

3. **Optical fiber power meter (10 μW – 1 mW, 632 nm-1650 nm) – secondary standard**

Our optical fiber power meter (OFPM) calibration system [24] is based on a pyroelectric radiometer. This consists of a pyroelectric detector with a highly absorbent and spectrally flat gold-black coating [55, 56]. The low reflectivity absorber allows the light to be input either through optical fiber (expanding beam) or as a collimated free-space beam. The fiber-coupled implementation allows calibration of fiber-coupled photodiode power meters, and the collimated beam configuration allows direct comparison to the LOCR.

The calibration with LOCR was carried out at a nominal power of 1 mW at wavelengths of 633 nm, 1319 nm, and 1550 nm using a direct substitution method [13]. The expanded uncertainty of the LOCR was 0.02 % and with the additional uncertainty sources of the OFPM and statistics of the measurement comparison, the OFPM calibration factor assigned by the LOCR has a relative expanded uncertainty ranging from 0.32 % to 0.44 % due to measurement conditions [24]. The OFPM is thus traceable through the LOCR.

4. **Single photon avalanche photodiode (0.5 fW – 400 fW, 851.7, 1533.6, 1550 nm) – secondary standard**

For power meter calibrations in the single-photon regime, we operate power meters with both free-space and fiber-optic coupling. We focus here on our single photon avalanche photodiode (SPAD) which is free-space coupled [26]. This meter is based on a silicon trap detector with a high-accuracy current-to-voltage amplifier and high accuracy voltmeter for readout.

This SPAD implementation is not intrinsically traceable to the SI and therefore requires a traceable calibration to one of our primary standards. We define the calibration factor for a single-photon detector as the detection efficiency (DE), which is equal to the ratio of the output signal (counts) and the input photon flux. An attenuated Ti:Sapphire laser source was used for the measurements. In the case of the SPAD, measurements were made at around 850 nm, the DE was determined at photon rates between 1500 counts per second (cps) and 1.6 million cps [26]. The input photon flux was determined through a calibration chain in which a silicon trap detector calibrated by the OFPM was used with calibrated attenuation of the photon flux to achieve the single photon level [26]. The OFPM is in turn traceable to the LOCR. This calibration chain of the free-space SPAD had a relative expanded uncertainty of 1.53 %, dominated by the spatial uniformity of the device under free-space coupling.

### 5. C-Series calorimeter (100 μW – 2 W, 325 nm – 2000 nm) – primary standard

For measurements in the range of 100 μW to 2 W, we use a vacuum enclosed isoperibol calorimeter denoted as "C-series" [14]. To report laser power, CW laser light is injected for an accurately-measured duration and the measured energy is divided by the injection duration time to yield power. The injected energy is absorbed by multiple internal reflections in a cupric-oxide coated cavity. The cavity is suspended by a weak thermal link, within an approximately constant-temperature (isoperibol) vacuum enclosure. The temperature of the absorbing cavity is continuously measured before, during, and after the injection to establish the temperature rise due to injected energy. Correcting for known thermal loss sources enables an estimate of the injected laser power [57].

The traceability of the C-series calorimeter has been established through electrical substitution [14]. In this approach an electrical heater (resistive wire wound around the absorber cavity) is used to inject a known amount of electrical power (traceable through a NIST-calibrated volt meter and a NIST-calibrated shunt resistor as well as the second) and compared to the calorimeter's calculated value of the injected power. The calibration factor of the C-series calorimeter was established in this manner. The relative expanded measurement uncertainty is typically 0.86 % [58].

### 6. Next-generation C-Series calorimeter (100 μW – 1 W, 325 nm – 2000 nm) – primary standard

Replacement technology for the C-series calorimeter is under development. This is denoted as the next-generation C-series calorimeter (Next Gen C) and is based on a planar absorbing bolometer [15]. This device uses a planar absorber with a coating of vertically-aligned carbon nanotubes for ultra-high absorptivity, obviating the need for an absorbing cavity. The temperature of the absorber is elevated to a constant value of 35 °C by electrical heating. Upon injection of laser power, the injected electrical energy is reduced to maintain the radiometer at constant temperature. The change in injected electrical power is determined by measuring the change in the current through and voltage across the heater. Traceability is through a NIST-calibrated voltmeter and shunt resistor and the second.

Preliminary results indicate a promising relative expanded uncertainty of 0.3 % [15]. This device has yet to be compared with our existing C-series calorimeter or other primary standards. However, upon verification of performance, we expect it will replace our current C-series instrument [58].

### 7. K-Series calorimeter (2 W – 10 kW, 800 nm to 10.6 μm) – primary standard

Using an isoperibol calorimeter approach [16] and precision injection timing to convert injected energy to power, we are able to measure injected laser power across the range from 2 W to approximately 150 W. As described in [14, 59] this is done with an absorbing multiple-bounce cavity, surrounded by a thermally insulating layer. We extend the operating range of the calorimeter with high-accuracy power attenuation. This is in the form of a reflective optical chopper [34] designed to be inserted in a high-power beam. The chopper blocks an accurately known fraction of light, attenuating the average power into the calorimeter. In this way, we enable an extended calorimeter operation range of 2 W to 10 kW.

Laser power measurement is by electrical substitution with traceability through a NIST-calibrated voltmeter and shut resistor and the injection time through the second. The relative expanded uncertainty is typically 1.1 % - 1.5 % [60].

### 8. Flowing water optical power meter (1 kW – 10 kW, 1070 nm and 10.6 μm) – primary standard

As CW laser power increases to a thousand watts and above, high accuracy power meters based on traditional calorimetry techniques become more difficult to use due to increased requirements for thermal management or precision attenuation. One approach for high-accuracy, high-power metrology is heat-balance calorimetry which we have implemented in the form of a flowing water optical power meter (FWOPM) [17]. The principle is that the laser light to be measured is injected onto a highly reflecting mirror inside a cavity. By spinning, the mirror redirects the light onto the absorbing walls of the highly absorbing optical cavity. The cavity is made of a water-cooled double copper shell blackened with a carbon nanotube coating to absorb the laser light. The flowing water not only extracts the injected heat, but its time-dependent temperature increase indicates the injected optical power. Laser power is measured from the difference between the output and input water temperature, the water's mass flow rate, and the heat capacity of the water.

In principle, the FWOPM can derive traceability from two different paths. With calibrated temperature sensors and mass flow meter as well as an accurate report of the water's specific heat capacity, the laser power can be determined. However, the FWOPM also has the capability to electrically heat the circulating water (thus emulating the laser-based heating). We use the latter electrical substitution technique as we can measure the injected electrical energy to a high degree of accuracy and by heating the FWOPM to the same temperature as was seen during laser injection, inequivalence due to uncompensated thermal loss mechanisms is negated. Traceability is through a NIST-calibrated voltmeter and shunt resistor yielding 1.6 % relative expanded uncertainty as reported in [17].

9. **Radiation pressure power meter (1 kW - 50 kW, 1070 nm) – primary standard**

Given the difficulty of measuring multikilowatt laser power levels with a thermal absorber, we have implemented a non-absorbing approach using radiation pressure. This alternative technique is based on measuring the force exerted by the laser beam as it reflects from a mirror. We describe such radiation-pressure based power measurement systems as radiation pressure power meters (RPPM) and, without the need to dissipate high powers within the measurement apparatus, power measurements are greatly simplified. Our radiation pressure power meter [20] establishes SI traceability of the force sensor through the kilogram in the form of NIST-calibrated test masses, the meter, and the second (since the mass measurement must be converted to a force using the acceleration of gravity [20]).

We have verified the stated uncertainty for the radiation pressure power meter through direct comparisons with the FWOPM and K-series calorimeter [61]. In both cases, the agreement between techniques was within the uncertainty of the measurements. The RPPM operates within our laboratory over the range of 1 kW to 10 kW (limited at the upper end only by the power presently available from our Yb-doped fiber laser source) with a relative expanded uncertainty of 1.6 %. However, we have performed careful measurements in other test environments with laser powers up to 50 kW [62] and found comparable uncertainties. We expect that power levels of 100 kW and above will also be measurable with comparable uncertainties (very preliminary work was carried out up to 92 kW [19]).

**Appendix B. Relating mass and optical power to the defining constants**

Here we support our assertions of Section IV regarding reduced uncertainty of radiation pressure power metrology by illustrating the relationship between mass, optical power, and the SI defining constants. We begin with the force induced by photon momentum (radiation pressure) and express results in terms of undefined unitless scaling factors (for simplicity) and SI defining constants. The force from radiation pressure as related to that from a test mass is

$$F = mg = \frac{PQ(\theta)}{c}. \tag{B1}$$

Here $c$ is the velocity of light, $g$ is local gravitational acceleration acting on mass, $m$, to produce a weight equivalent to the photon pressure force, $F$, and $P$ is the optical power of the light. $Q(\theta)$ is a factor describing the effects of reflection, absorption and scattering processes on $F$ from light incident on a surface at angle $\theta$. This factor is 2 for perfect specular reflection at normal incidence, or $\cos(\theta)$ for total absorption at incidence angle $\theta$, for example.

Consider the power in terms of the individual photons reflecting from the surface. In this case

$$P = h\nu_p r, \tag{B2}$$

where $h$ is Planck's constant, $\nu_p$ is the optical frequency, and $r$ is the photon arrival rate. We note that for a photon-on-demand system, this would be a constant rate, but is more commonly an average rate obeying Poisson statistics. Both $\nu_p$ and $r$ can be quantified relative to the hyperfine splitting frequency of cesium 133, $\Delta\nu(^{133}\text{Cs})_{\text{hfs}}$, such that

$$\nu_p = A\Delta\nu(^{133}\text{Cs})_{\text{hfs}},$$

$$r = B\Delta\nu(^{133}\text{Cs})_{\text{hfs}}$$

using scaling constants $A$ and $B$, respectively. Local gravitational acceleration can also be related to the fundamental constants using length, $l_g$, and time, $t_g$, such that

$$g = \frac{l_g}{t_g^2} = \frac{cD\Delta\nu(^{133}\text{Cs})_{\text{hfs}}}{E^2} \tag{B3}$$

using scaling constants $D$ and $E$ for the measurements of $l_g$ and $t_g$, in terms of the $c$ and $\Delta\nu(^{133}\text{Cs})_{\text{hfs}}$. Using Equation (B1) and Equation (B3),

$$m = \frac{h\Delta\nu(^{133}\text{Cs})_{\text{hfs}}}{c^2} KQ(\theta), \tag{B4}$$

where the dimensionless scaling constant $K=ABE^2/D$. This is directly analogous to the fundamental relation used in the realization of mass in the Kibble balance [63] and electrostatic force balance [51], in which

$$m = \frac{h\Delta\nu(^{133}\text{Cs})_{\text{hfs}}}{c^2} L. \tag{B5}$$

Here, the scaling constants used to obtain mass from electrical, dimensional and frequency metrology can be collected in a single term, $L$. This demonstrates photon counting power measurements are compatible with the SI redefinition through the kilogram.

Similarly, the measurement of classical radiation pressure force is a realization of photonic power if it is established to be equal to the force directly generated by a Kibble balance or electrostatic force balance. It is SI-traceable if a mass traceable to one of these balances is used as a reference for the photon pressure force measurement. Using Equations (B1), (B3), and (B5), we obtain

$$P = h\Delta\nu(^{133}\text{Cs})_{hfs}^2 \frac{LD}{Q(\theta)E^2}, \tag{B6}$$

where the only fundamental constants required are Planck's constant and the cesium 133 hyperfine splitting frequency. This illustrates the possibility that the optical Watt can be obtained as a force measurement from these two fundamental constants without need for the kilogram within the SI.

**Appendix C. Foundation for uncertainty statements**

This appendix describes the approach taken to establish the uncertainty expressions used in the main body of the paper. In particular, expressions for three quantities and their associated uncertainties are derived – the estimated power $\hat{\mathcal{P}}_a$ and its relative expanded uncertainty $U_{\hat{\mathcal{P}}_a}$ when the power meter is a primary standard, the comparison factor $K_{a,b}$ and its relative expanded uncertainty $U_{K_{a,b}}$, and finally estimated power $\hat{\mathcal{P}}_{a \to b}$ and the relative expanded uncertainty $U_{\hat{\mathcal{P}}_{a \to b}}$ for power measured by power meter "$a$" whose traceability is through power meter "$b$" (a primary standard).

We begin with laser power meter "$a$" on which is incident a laser power $P_0$. If the power reported by power meter "$a$" is $P_a(P_0)$, then the best estimate $\hat{\mathcal{P}}_a$ of the actual power can be expressed as

$$\hat{\mathcal{P}}_a = P_a(P_0) \cdot z_a, \tag{C1}$$

where $z_a$ represents systematic error sources. Specifically, $P_a(P_0)$ is the final value reported by the power meter and has been corrected for any known systematic biases. Therefore, to the best of our knowledge, $z_a = 1$. The distinction between $\hat{\mathcal{P}}_a$ and $P_a(P_0)$ is merely a formalism to allow a separation of uncertainty sources. The uncertainty of $P_a(P_0)$ is based on repeated measurements and purely statistical and commonly referred to as Type A [27]. The uncertainty of the $z_a$ term is purely non-statistical (systematic) and commonly referred to as Type B [27]. We express the relative expanded uncertainty $U_{\hat{\mathcal{P}}_a}$ of the estimated power as

$$U_{\hat{\mathcal{P}}_a} = 2u_{\hat{\mathcal{P}}_a} = 2\frac{u(\hat{\mathcal{P}}_a(P_0))}{\hat{\mathcal{P}}_a(P_0)} = 2\sqrt{\left(\frac{u(P_a(P_0))}{P_a(P_0)}\right)^2 + \left(\frac{u(z_a)}{z_a}\right)^2}, \tag{C2}$$

where the coverage factor $k=2$ defines an interval having a level of confidence of approximately 95 %. To simplify the notation, we name the relative uncertainty expressions according to their statistical nature so that Equation (C2) becomes

$$U_{\hat{\mathcal{P}}_a} = 2\sqrt{u_{a,stat}^2 + u_{a,non-stat}^2}. \tag{C3}$$

The power and uncertainty expressions for power meter "$b$" are analogous to those for power meter "$a$".

The comparison factor,

$$K_{a,b}(P_0) = \frac{P_a(P_0)}{P_b(P_0)}, \tag{C4}$$

expresses the ratio between the power reported by power meter "$a$" and that reported by power meter "$b$" when measuring the same power $P_0$. Since this comparison factor represents merely the ratio of reported powers, it does not indicate the accuracy of either power meter. Since $P_a$ and $P_b$ have only statistical uncertainties, the relative expanded uncertainty of the comparison factor is given as

$$U_{K_{a,b}} = 2u_{K_{a,b}} = 2\frac{u(K_{a,b}(P_0))}{K_{a,b}(P_0)} = 2\sqrt{\left(\frac{u(P_a(P_0))}{P_a(P_0)}\right)^2 + \left(\frac{u(P_b(P_0))}{P_b(P_0)}\right)^2} = 2\sqrt{u_{a,stat}^2 + u_{b,stat}^2}. \tag{C5}$$

As described in the main body of the paper, the comparison factor for a power meters "$a$" and "$b$" can be found even when they operate at different powers and or wavelengths. This requires comparison to an intermediate (transfer) power meter "$x$" with a spectral responsivity ratio $\eta_x$ and a power nonlinearity factor $\gamma_x$ described by Equations (8) through (10). This mediated comparison factor

$$K_{a,b,x} = \frac{K_{x,b}(P_2,\lambda_2)}{K_{x,a}(P_1,\lambda_1)} \gamma_x \eta_x, \tag{C6}$$

has a relative expanded uncertainty

$$U_{K_{a,b,x}} = 2u_{K_{a,b,x}} = 2\frac{u(K_{a,b,x})}{K_{a,b,x}} = 2\sqrt{u_{K_{x,b}}^2 + u_{K_{x,a}}^2 + \left(\frac{u(\gamma_x)}{\gamma_x}\right)^2 + \left(\frac{u(\eta_x)}{\eta_x}\right)^2}. \tag{C7}$$

The transitive relationship of Equation (C6) allows any two power meters to be virtually compared if there is an unbroken chain of comparisons between them,

$$K_{n,1} = K_{n,n-1}K_{n-1,n-2}\ldots K_{3,2}K_{2,1} = \prod_{i=1}^{n-1} K_{i+1,i}. \tag{C8}$$

Since each of the *K* values in Equation (C8) are measured independently of the others their errors are uncorrelated and their covariance is zero. Therefore, the relative expanded uncertainty of the net comparison factor (e.g. $K_{n,1}$ in Equation (C8)) comes from the quadrature sum of the relative standard uncertainties of each of the direct comparisons (GUM 5.1.6) [27].

$$U_{K_{n,1}} = 2u_{K_{n,1}} = 2\frac{u(K_{n,1})}{K_{n,1}} = 2\sqrt{u_{K_{n,n-1}}^2 + u_{K_{n-1,n-2}}^2 + \cdots + u_{K_{2,1}}^2} = 2\sqrt{\sum_{i=1}^{n-1} u_{K_{i+1,i}}^2}, \quad (C9)$$

where the individual *K* values and their relative uncertainties can be either the direct (Equations (C4) and (C5)) or the mediated (Equations (C6) and (C7)) comparison factors.

Finally, we can use the power reported by power meter "*a*" to produce an estimated power value referenced to power meter "*b*". That is, the power estimated by power meter "b" is

$$\hat{\mathcal{P}}_b = P_b(P_0) \cdot z_b. \quad (C10)$$

Combining this with the comparison factor between power meters "*a*" and "*b*" (Equation (C4)) gives

$$\hat{\mathcal{P}}_b = \frac{P_a(P_0)}{K_{a,b}(P_0)} \cdot z_b, \quad (C11)$$

which is the power estimate from a measurement by power meter "*a*" whose traceability is through power meter "*b*". The resulting relative expanded uncertainty is found from Equations (C2) through (C5) to be expressible as

$$U_{\hat{\mathcal{P}}_{a \to b}} = 2u_{\hat{\mathcal{P}}_{a \to b}} = 2\sqrt{u_{a,stat}^2 + u_{Ka,b}^2 + u_{b,non-stat}^2}. \quad (C12)$$

Here, as in all cases above, the capital *U* represents the relative expanded uncertainty, where the coverage factor *k*=2 defines an interval having a level of confidence of approximately 95 %.